# An Evaluation of Link Performance Based on Rainfall Attenuation for a LEO Communication Satellite Constellation Over Africa

Osoro O.B, *George Mason University,* and Suki D.S, *University of Strathclyde, Glasgow*

Corresponding Author Address: bosoro@gmu.edu

**Abstract-***The emergence of Low Earth Orbit (LEO) satellites is seen as a solution to providing affordable broadband internet in unplanned and sparsely populated areas and supporting the 5G rollout. However, little research has been done on viability of communicating with ground stations especially in unplanned and sparsely populated areas such as Africa where they are applicable in narrowing the broadband gap. Coincidently, these regions are in tropical and equatorial climatic zones where rainfall attenuation on microwave frequency links is a major problem. The current ITU-R p.618-8 model which relies on rain intensity values from ITU-R P.837-7 is the standard way of quantifying rain attenuation. Since this model was based on data from temperate climatic regions, they overestimate attenuation leading to system overdesign. Furthermore, rain gauge measurements are sometimes unavailable in potential gateway stations for such regions in Africa, thus calling for experimental setup that might take more than four years to accurately model the attenuation* [1] [2]*. Nevertheless, precipitation data from various global satellite missions are always available for such regions.*

*In this paper, we apply the ITU-R P.618-8 model with data from the ITU-R P.837-7, Tropical Rain Measuring Mission (TRMM) and Global Precipitation Mission (GPM) satellite to determine the level of attenuation and available link margin for a LEO system such as Telesat. The specific and predicted attenuation for chosen six candidate ground stations (Abuja, Hartbeesthoek, Cairo, Longonot, Port Louis and Praia) is computed and results presented for 0.001%-1% unavailability time in a year. Setting a link margin of 0.36dB, the available link margin and the best candidate ground station for a LEO system such as Telesat is determined. The approach used can be implemented for other potential ground stations and LEO communication systems over Africa.*

## I. INTRODUCTION

Low Earth Orbit (LEO) satellites have attracted massive attention recently due to their technical advantages in terms of round-trip time (RTT) and large available bandwidth [3]. The location of the satellites in lower altitudes (500km-2000km) allows the use of high frequency bands such as Ka, Ku, Q and V bands without limiting the range of transmission as geosynchronous earth orbit (GEO) satellites [4]. Consequently, they are now seen as an alternative for augmenting terrestrial New Radio (NR) technologies such as 5G/6G and other emerging applications such as Internet of Things (IoT), ultra-reliable low latency communications (URLLC) and enhanced mobile broadband (eMBB) [4] [5].

Unlike optical fiber, LEO satellites can be employed to provide and narrow the "broadband gap" in sparsely populated areas, regions with extreme topographies such as cliffs, valleys and steep slopes as established by [6]. With majority of sub-Saharan African countries characterized by unplanned and sparse rural settlement, LEO satellites offer a cheaper, scalable and alternative way of providing broadband connectivity as demonstrated in case studies of Malawi, Kenya, Senegal and Uganda [6]. Unfortunately, these regions fall within tropical or equatorial climate characterized by heavy and peculiar rainfall patterns [1]. This has a grave impact on the performance, link availability and general Quality of Service (QoS) of LEO communication satellites due to high frequencies used that are susceptible to rainfall attenuation [7]. The recommended International Telecommunication Union (ITU-R) models for characterizing rainfall attenuation works best in temperate regions as opposed to equatorial and tropics where severe degradation starts as low as 7GHz [8] [9].

Therefore, this paper analyzes the uplink performance of a working LEO communication constellation

(Telesat) in regards to six candidate ground stations across Africa (Abuja 9.0108°N 7.2714°E, Hartbeesthoek 25.8889°S 27.6853°E, Cairo 29.9675°N 31.2750°E, Longonot 1.0178°S 36.4969°E, Port Louis 20.1389°S 57.7253°E, Praia 15.1061°N 23.5131°W) by only considering rainfall attenuation and using the ITU-R P.618-8 model. Rain rate measurements from the NASA's Global Precipitation Measurement (GPM) satellites are compared with values based on the ITU-R P.837-7 to accurately quantify the rainfall attenuation in the six candidate ground stations. We then proceed to calculate the Carrier to Noise Ratio (CNR) and link availability to establish QoS of Telesat in the proposed candidate ground stations.

I. LITERATURE REVIEW

Attempts to accurately characterize rainfall attenuation on earth-space communication over Africa and generally in tropical and equatorial regions is not new. Oje, Ajewole and Sarkar (2008), quantified the rainfall attenuation at Ku and Ka band in Nigeria for its geostationary satellite (NIGCOMSAT-1). The rain rate data developed using Moupfouma tropical model for the GEO satellite, NIGCOMSAT-1 yielded more accurate rain attenuation results compared to ITU. A similar approach has been adopted by Djuma, et al., (2016) to analyze the difference in attenuation level for eight Rwandan cities using the Chebil model. Apart from overestimation by ITU-R model, the results showed that cities located in the western region with high intensity rainfall experienced more attenuation as compared to those in the East [10]. Despite the improvement in quantifying the rainfall attenuation in these two tropical regions, there is need to consider LEO satellites as opposed to GEO as well as integrate satellite data that can be applied over wide regions where station rain gauge data is not available.

Yusuf, et al., (2016) empirically quantified the rainfall attenuation level for Ku (15GHz) to determine the overestimation magnitude of ITU-R model for a GEO satellite. The four-year results confirmed that ITU-R model did not accurately represent the rain attenuation for a tropical region such as Malaysia [11]. However, this did not consider an earth-space communication but a terrestrial link. Shrestha and Chois (2017), extended the work of [11] by characterizing the degree of rainfall attenuation in communicating with its two GEO satellites, Koreasat 6 and COMS1. Although the study focused on deriving the power coefficients for computing specific attenuation in Ku (12.25GHz) and Ka (19.8GHz & 20.73GHz) bands, the results demonstrate the importance of accurately quantifying attenuation in tropical regions such as South Korea. A detailed comparison with six prediction models (Unified, Dissanayake Allnutt and Haidara, Simple Attenuation Model, Crane Global, Ramachandran and Kumar Model) further revealed the poor performance of ITU-R model [12].

While the approaches in [8] [1] [2] [9] [13] [11] are based on rain gauge data from experimental setup, Rimven, et al., (2018) used Tropical Rainfall Measuring Mission (TRMM) and UK rain radar measurements from NIMROD to develop a more accurate method for quantifying rainfall attenuation rate in tropical regions such as Africa [14]. This opens an avenue for exploring the approach with other satellite data such as the GPM data to compute the level of attenuation on links for LEO satellites over tropical regions such as Africa. Furthermore, the results from such an approach can be used to calculate the CNR and link availability of new LEO communication constellations such as Telesat, over the same region.

*A. Research Objectives*

The objective of this paper is to compute rainfall attenuation using ITU-R P.618-8 with rain rate data from TRMM and GPM for the six candidate ground stations. The uplink performance and availability for each of the candidate ground stations is established for existing LEO communication constellation, Telesat. From the results, the best candidate gateway station for a system such as Telesat is established.

1. Telesat System Uplink Parameters

Table 1 outlines the system parameters for Telesat from Federal Communications Communication (FCC), literature and press briefing as of 2018 [15].

| Parameter | Value | Unit |
|---|---|---|
| Frequency | 28.5 | GHz |
| Bandwidth | 2.1 | GHz |
| Transmit Antennae Diameter | 3.5 | m |
| EIRP | 75.9 | dBW |
| Elevation Angle | 20 | deg |
| Free Space Path Loss | 189.3 | dB |
| Receiver Antenna Gain | 31.8 | dBi |
| System Temperature | 868.4 | K |
| Link Margin | 0.36 | dB |
| Altitude | 1200 | km |

*Table 1 Beam link budget for uplink gateway of Telesat Adopted from Portilo et al (2019)*

II. METHODOLOGY

Six candidate ground stations based on three conditions as suggested by [16] are selected across Africa and their geographical coordinates determined from Google Earth application. The conditions are:

i. Global coverage: Ability of the ground station to serve the whole of Africa including

the countries in the middle of the oceans such as Mauritius, Madagascar and Cape Verde.
ii. A 2000km minimum distance of separation between any two ground stations to ensure that the weather conditions are not spatially correlated.
iii. Realistic or potential location characterized by some space activities.

The selected ground stations are listed in and shown in **Error! Reference source not found.**.

| Country | Location | Latitude | Longitude | Height Above the Sea Level (m) |
|---|---|---|---|---|
| Nigeria | Abuja | 9.010833°N | 7.271389°E | 348.00 |
| South Africa | Hartbeesthoek | 25.88889°S | 27.68528°E | 1385.0 |
| Egypt | Cairo | 29.96750°N | 31.27500°E | 40.000 |
| Kenya | Longonot | 1.017778°S | 36.49694°E | 1715.0 |
| Mauritius | Port Louis | 20.13889°S | 57.72528°E | 29.000 |
| Cape Verde | Praia | 15.10611°N | 23.51306°W | 84.000 |

*Table 2 Selected Candidate ground stations in Africa for Telesat uplink performance analysis.*

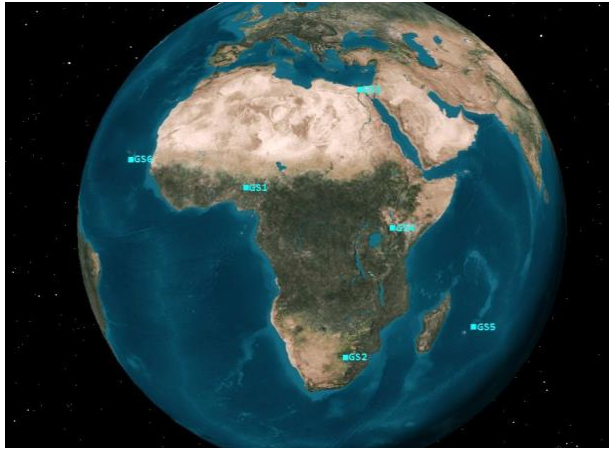

*Figure 1 World Globe showing the distribution of the candidate ground stations labelled GS1 to GS6 respectively according to the order in Table 2 designed using Standard Tool Kit (STK).*

### A. Data

The integrated multi-satellite retrievals for TRMM and GPM IMERG rain intensity data for the six-candidate ground station are downloaded from NASA's website. The TRMM data was from January 1/1/2005 to 1/1/2014 while the GPM covered January 2015 to June 2020. The passive microwave (PMW) sensors hosted in LEO satellites are the principal component of the IMERG dataset [17]. Since the measurement are sometimes sparse, GEO Infrared (IR) satellite estimates and ground precipitation gauge analyses are integrated to improve the accuracy [17]. The GPM data are more accurate than its predecessor, TRMM due to three improvements. 1) an increase in inclination from 35⁰ to 65⁰ thus guaranteeing coverage of high latitude climatic zones and reduction in amount of radiometer sampling [17]. 2) addition of two radar frequencies hence more sensitivity to light precipitation 3) the 165.5GHz and 183.3GHz channels provide key solid and light precipitation sensing information [17].

### B. Estimation of Predicted Rain Attenuation

The computation of rain attenuation using ITU-R documentation is a 10-step process. The ITU-R P.618-8 is the latest version of the model for quantifying the rainfall attenuation in microwave satellite links [18].

The fourth step of the procedure requires rainfall intensity rate exceeded for 0.01% of an average year data to calculate specific attenuation using Equation 1.

$$\Upsilon R(dB/km) = \kappa R_{0.01}^{\alpha} \qquad (1)$$

Where $\kappa$ and $\alpha$ are coefficients that are a function of Telesat's uplink 28.5GHz frequency that were obtained from ITU-R P.838-3 by integration [19]. The polarization is assumed to be vertical since it limits attenuation by rain drops [10].

The predicted attenuation above 0.01% for the year was calculated using.

$$A_{0.01} = \Upsilon R L_e (dB) \qquad (2)$$

$$\frac{C}{N} = EIRP(dBW) + \frac{G_r}{T}(dBi/K) - FSPL(dB) - [R_{Att} + OT_{Loss}](dB) - 10.\log_{10}(k.T.B) \quad (4)$$

Where $L_e$ is effective path length calculated as

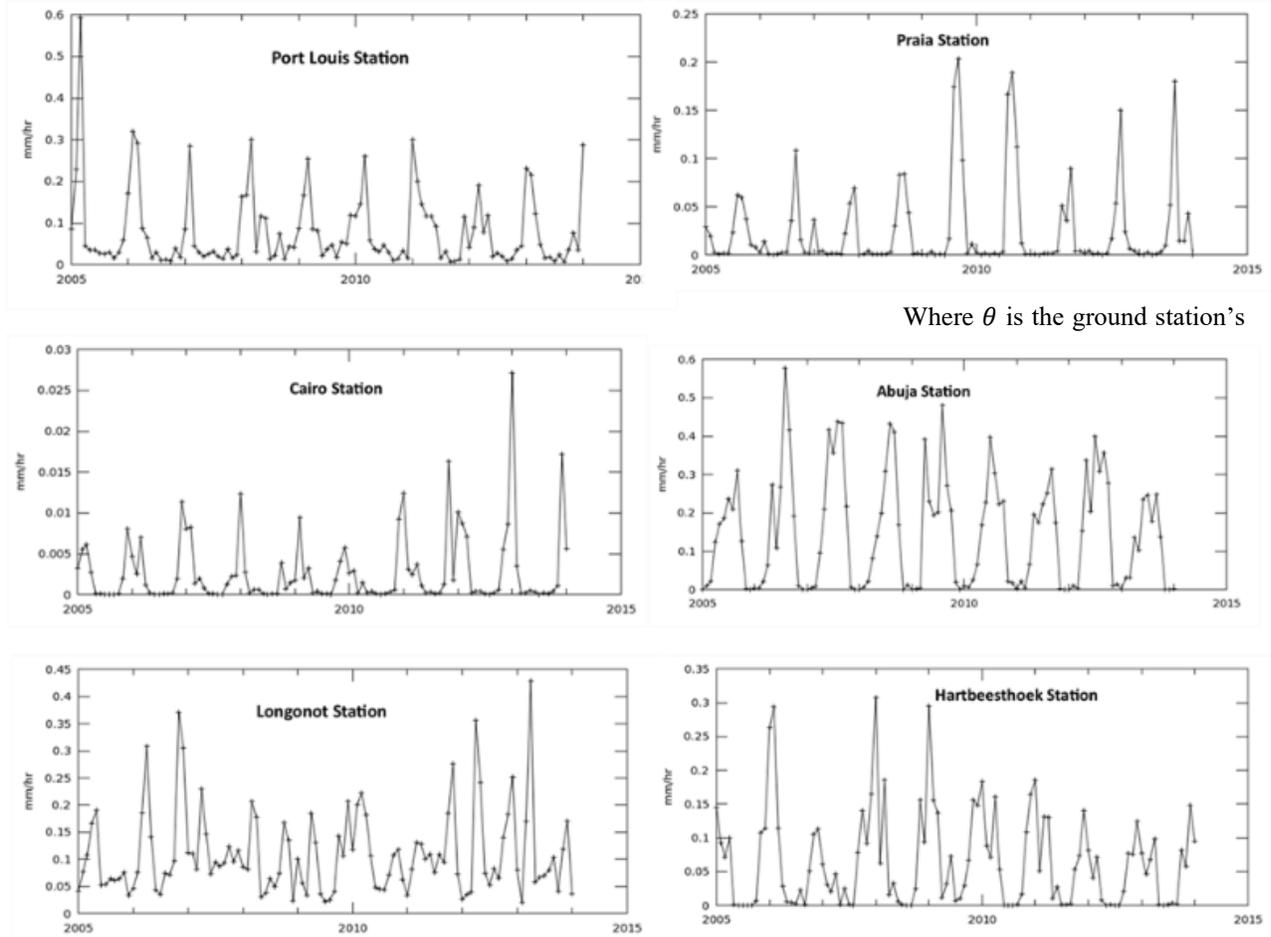

Where $\theta$ is the ground station's

Figure 2 Time series, area-averaged of precipitation rate monthly 0.25degrees (TRMM 3B43 V7 data) in mm/hr between January 2005 to January 2014 for Port Louis, Praia, Cairo, Abuja, Longonot and Hartbeesthoek proposed candidate Stations.

explained in ITU-R P.618-8 documentation [18]. The predicted attenuation rate for other percentages exceeded in the range 0.001%-1.0% was then calculated for the six ground stations using.

absolute latitude and $\varepsilon_{min}$ elevation angle. The resulting rainfall attenuation rate are then used in computation of the CNR to establish the effect of train, available link and the best performing ground station

$$A_p = A_{0.01}\left(\frac{p}{0.01}\right)^{-[0.655+0.033\ln(p)-0.045\ln(A_{0.01})-z\sin\varepsilon_{min}(1-p)]} \quad (3)$$

Where $p$ is the percentage probability under investigation and z is a magnitude, whose value depends on the station latitude [18]:

using the Friis Transmission Equation (Equation 4) [20].

$$For\ p \geq 1\%,\ z = 0\ and\ p < 1\%\ z = 0\ for\ |\theta| \geq 36°$$

$$z = -0.005(|\theta| - 36)\ for\ \varepsilon_{min} \geq 25°\ and\ |\theta| < 36°$$

$$z = -0.005(|\theta| - 36) + 1.8 - 4.25\sin\varepsilon_{min}, for\ \varepsilon_{min} < 25°\ and\ |\theta| < 36°$$

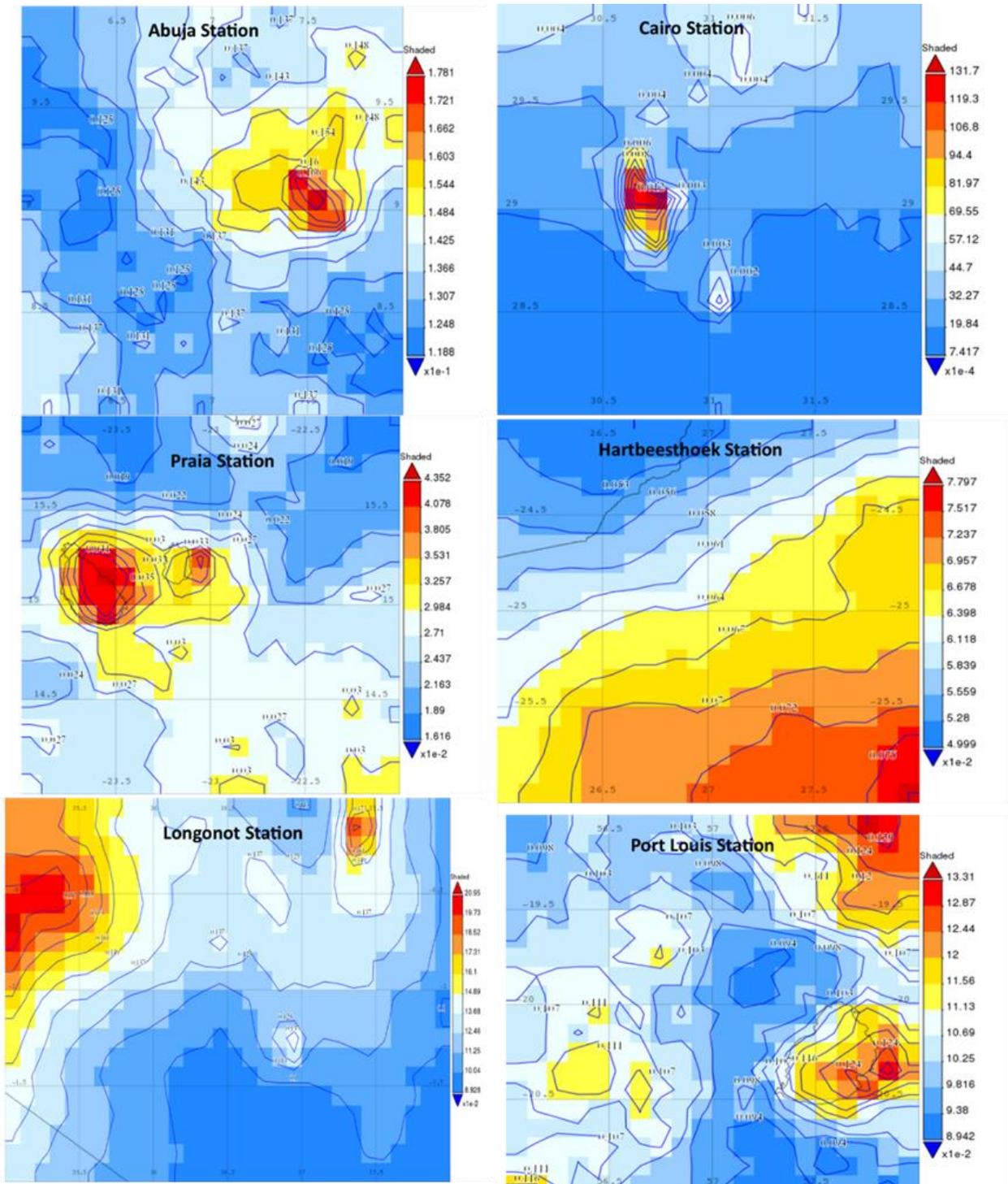

*Figure 3 Contour for time-averaged overlay map of merged satellite-gauge precipitation (3IMERGM v06) in mm/hr monthly 0.1 degrees spatial resolution from January 2015 to June 2020 for the six candidate stations*

Where EIRP is effective isotopic radiated power of the ground station antenna, $G_r$ is receiver gain, T system temperature, FSPL, free space path loss, $OT_{Loss}$, sum of all other losses, k Boltzmann constant, B bandwidth and $R_{Att}$, the rain attenuation obtained from Equation 3.

## III. RESULTS

First, the rain intensity data from ITU-R P.837-7 for each of the stations are used to compute the rain attenuation for each of the candidate stations. The results act as the baseline for comparing expected attenuation for the stations.

Next, the TRMM data were downloaded and analyzed using NASA's online software, Giovanni to retrieve the rain rates. The plotted results for the TRMM data of all the six candidate stations is shown in **Error! Reference source not found.**.

From the plots, Hartbeesthoek, Longonot and Abuja stations records the highest rain intensity with an average of 0.0657, 0.1105 and 0.1455mm/hr. Generally, Longonot, Abuja and Hartbeesthoek falls within the tropical climatic zones that experience high of the defined geographical boundary experience high rain intensity as indicated by the red and yellow contours. Equally, the contour overlay maps of Port Louis and Praia indicate lower rain intensity levels as indicated previously by the time series data from TRMM in **Error! Reference source not found.**.

The ITU-R P.837-7, TRMM and GPM results are then used in the subsequent calculations.

Using the rain intensity values from ITU-R P.837-7, the average rain rate, specific, rainfall attenuation and CNR using equations (1), (3) and (4) respectively for 0.01% unavailability time are computed. The CNR results are shown in **Error! Reference source not found.**.

Hartbeesthoek and Longonot stations have the highest altitudes (1385m) and (1715m) while equally

| Ground Station | Location | Rain Attenuation (dB) | CNR (dB) | Margin (dB) | Available Margin (dB) |
|---|---|---|---|---|---|
| **Abuja** | 9.010833°N, 7.271389°E, 348.00 | 34.1808 | -31.1143 | 0.36 | -31.4743 |
| **Hartbeesthoek** | 25.88889°S, 27.68528°E, 1385.0 | 49.0126 | -45.9461 | 0.36 | -46.3061 |
| **Cairo** | 29.96750°N, 31.27500°E, 40.0m | 31.6560 | -28.5895 | 0.36 | -28.9498 |
| **Longonot** | 1.017778°S, 36.4969°E, 1715.0m | 40.2605 | -37.194 | 0.36 | -37.554 |
| **Port Louis** | 20.13889°S, 57.72528°E, 29.0m | 27.5556 | -24.4891 | 0.36 | -24.8491 |
| **Praia** | 15.10611°N, 23.51306°W, 84.0m | 28.0972 | -25.0307 | 0.36 | -25.3907 |

*Table 3 Results of the rainfall attenuation, CNR and available link margin based on ITU-R P.837-7 model for all the six candidate ground stations for 0.01% unavailability time*

peculiar rainfall patterns [1]. Cairo being in a desert and temperate region records the least rain intensity. Similarly, Praia records least rain intensity while Port Louis records average but consistent rainfall pattern for the period under study.

The GPM data is then analyzed to cover for the period between January 2015 and June 2020. These were then analyzed using NASA' online software, Giovanni to generate the rain rate for each of the six candidate ground stations. The contour for time-averaged rain rate overlay map for each of the stations is shown in **Error! Reference source not found.**.

recording the highest rainfall attenuation rate. Whether the altitude contributes to the high attenuation rate is beyond the scope of this study. The correlation can only be established by simulating or experimenting with regions of different altitude but with all other parameters remaining constant to ascertain it. Literature [21] has already pointed to the relationship between latitude and rainfall attenuation but not altitude. The least attenuation levels are recorded at Port Louis, Praia and Cairo thus leading to lower CNR. However, the results indicate that none of the stations could have their link close for the 0.01% unavailability time.

| Ground Station | Rainfall Attenuation (dB) | CNR (dB) | Margin (dB) | Available Margin (dB) |
|---|---|---|---|---|
| **Abuja** | 10.5587 | -7.4922 | 0.36 | -7.8522 |
| **Hartbeesthoek** | 22.7269 | -19.6604 | 0.36 | -20.0204 |
| **Cairo** | -7.8440 | 10.9105 | 0.36 | 10.5505 |
| **Longonot** | 15.5252 | -12.4587 | 0.36 | -12.8187 |
| **Port Louis** | -0.3620 | 3.4285 | 0.36 | 3.0685 |
| **Praia** | -1.7871 | 4.8536 | 0.36 | 4.4936 |

*Table 4 Results of the rainfall attenuation, CNR and available link margin based on GPM data for all the six candidate ground stations for 0.01% unavailability time*

The contour for time-averaged overlay maps for the six candidate stations conform to the previous results obtained from TRMM. The areas bound by the geographical extent of the ground stations generally record high rain intensity for Hartbeesthoek, Longonot and Abuja. However, for Cairo, only a small portion

Equally, for the TRMM data, the average rain rate is used to compute the specific, rainfall attenuation and CNR using equations (1), (3) and (4) respectively for 0.01% unavailability time. The CNR results are shown in **Error! Reference source not found.**.

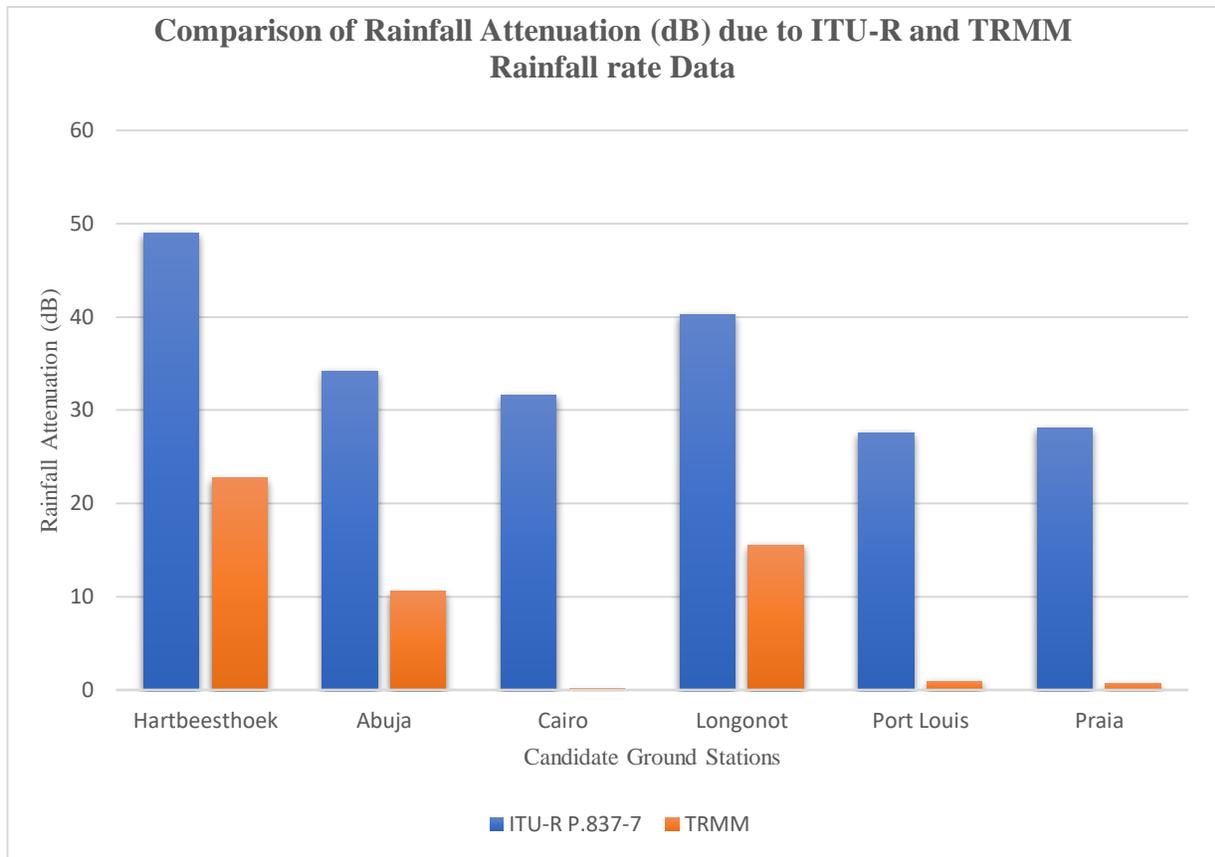

*Figure 4 An illustrative comparison of rainfall attenuation results due to rainfall rate values from ITU model and TRMM satellite data for 0.01% unavailability time.*

Abuja, Longonot and Hartbeesthoek records the highest attenuation rates of ≈ 11, 16 and ≈ 23 $dB$. Cairo records the least attenuation of (-7.844dB) followed by Praia (-1.7871dB) and Port Louis (-0.362dB). The rain attenuation results are directly proportional to the rain rate values for the regions obtained from the TRMM data. The CNR results also reveals that only Cairo, Praia and Port Louis link could close for the 0.01% unavailability time.

Just like with ITU-R P.837-7 and TRMM, the GPM rain rate results from **Error! Reference source not found.** are imputed in equation (1) to calculate the specific and resulting rainfall attenuation for the ground stations for 0.01% unavailability time. Using the rain attenuation values and equation (4), the CNR for each of the stations is obtained. With a margin (M) of 0.36dB for Telesat, the available link for each of the stations is computed. **Error! Reference source not found.** shows these results.

Cairo, Praia and Port Louis records the least attenuation of -13.2802dB, -1.3956dB and 0.7753dB, respectively. This is contrary to Abuja, Longonot and Hartbeesthoek that records the highest attenuation of 10.3059dB, 16.3896dB and 22.7947dB, respectively. Based on the 0.36dB link margin, Abuja, Longonot and Hartbeesthoek fall below the link margin.

We then compared the attenuation results from the three sources (ITU-R P.837-7, TRMM and GPM) to determine the best values to use for the final link margin calculation of the proposed candidate stations. First, the difference between the ITU-R P.837-7 and TRMM rain attenuation values and the overestimation percentage are calculated as illustrated in

| Ground Station | Rainfall Attenuation (dB) | CNR (dB) | Margin (dB) | Available Margin (dB) |
|---|---|---|---|---|
| **Abuja** | 10.3059 | -7.2394 | 0.36 | -7.5994 |
| **Hartbeesthoek** | 22.7947 | -19.7282 | 0.36 | -20.0882 |
| **Cairo** | -13.2802 | 22.5073 | 0.36 | 22.1473 |
| **Longonot** | 16.3896 | -13.3231 | 0.36 | -13.6831 |
| **Port Louis** | -0.7753 | 3.8418 | 0.36 | 3.4818 |
| **Praia** | -1.3956 | 4.4621 | 0.36 | 4.1021 |

Table .

| Station | Overestimation Percentage |
|---|---|
| Abuja | 70% |
| Hartbeesthoek | 54% |
| Cairo | 142% |
| Longonot | 59% |
| Port Louis | 97% |
| Praia | 105% |

*Table 6 Overestimation percentage magnitude of ITU-R P.618-8 model compared to GPM data*

| Station | Overestimation Percentage |
|---|---|
| Hartbeesthoek | 54% |
| Abuja | 69% |
| Cairo | 125% |
| Longonot | 61% |
| Port Louis | 101% |
| Praia | 106% |

*Table 5 The magnitude of overestimated values.*

| Station | Overestimation Percentage |
|---|---|
| Abuja | 70% |
| Hartbeesthoek | 54% |
| Cairo | 142% |
| Longonot | 59% |
| Port Louis | 97% |
| Praia | 105% |

*Table 6 Overestimation percentage magnitude of ITU-R P.618-8 model compared to GPM data*

Generally, the attenuation results from ITU-R P.837-7 for the ground stations is high as depicted by the overestimation percentages. **Error! Reference source not found.** shows the difference graphically.

The highest overestimation has been recorded for Cairo, Praia and Port Louis ground stations. Coincidentally, these are the stations that recorded least rain rate as shown in the overlay contour maps. This implies that ITU-R P.837-7 also performs worse for regions with low rain intensity.

Since the TRMM data was for the period 2005-2014, we also compared ITU-R P.837-7 results with the updated GPM data of 2015-2020 to establish the degree of overestimation/underestimation.

Like TRMM, GPM results showed that the ITU-R P.837-7 overestimates the rain attenuation for the candidate states. The overestimation percentage for Praia, Cairo and Port Louis is 105%,142% and 97% respectively. This is not the case with regions with high rain rates such as Hartbeesthoek, Longonot and

Abuja with values of 54%, 59% and 70% respectively. **Error! Reference source not found.** graphically illustrates the overestimation of rainfall attenuation based on the rain rate data provided by the model, TRMM and GPM. Finally, we compared the overestimation/underestimation percentage of the two missions (TRMM and GPM) to determine the best set of results to use for the final link margin computations. The results are shown in Table 5 Estimation comparison of TRMM and GPM results for the six candidate ground stations.

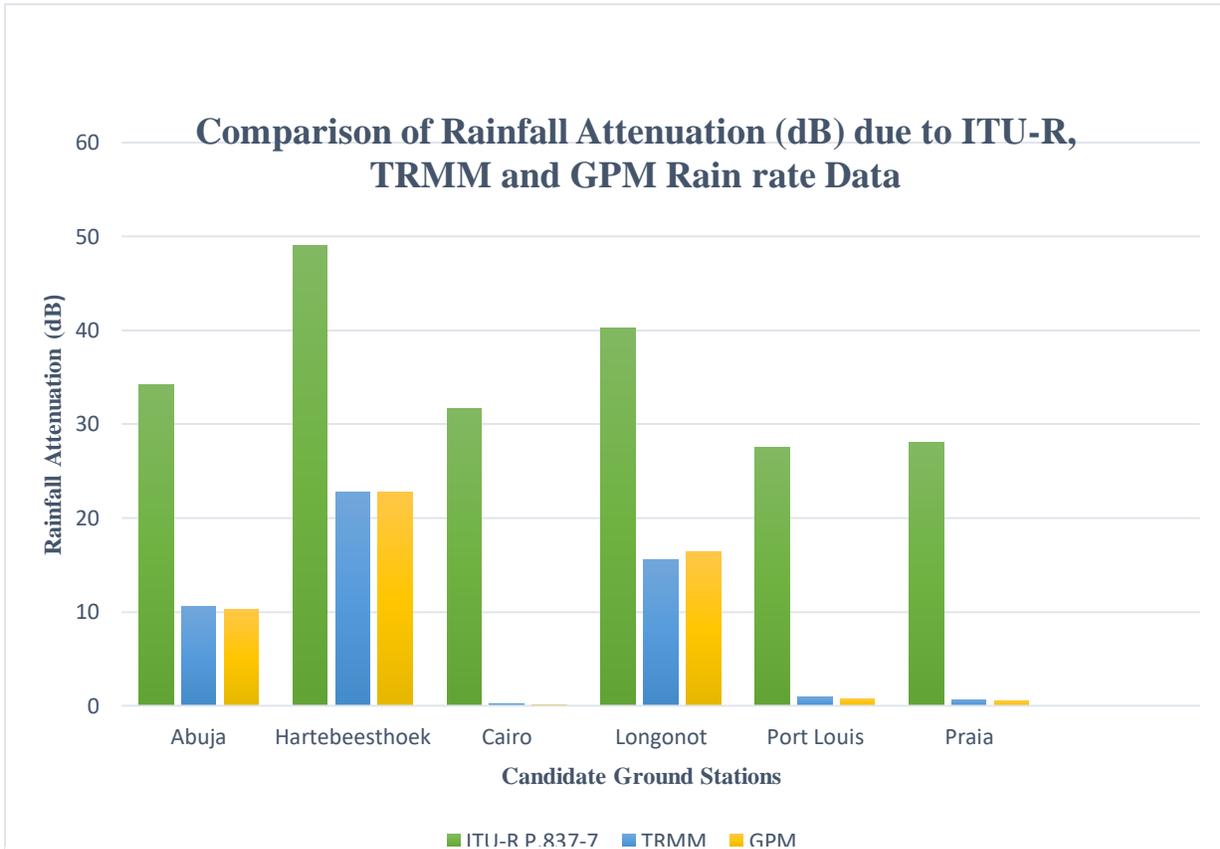

*Figure 5 A graphical comparison of the rainfall attenuation rates of the candidate stations based on rain rate values from ITU model, TRMM and GPM satellite data*

| Ground Station | Rainfall Attenuation (dB) | CNR (dB) | Margin (dB) | Available Margin (dB) |
|---|---|---|---|---|
| Abuja | -24.1363 | 29.4026 | 0.36 | 29.0426 |
| Hartbeesthoek | -13.5125 | 18.7788 | 0.36 | 18.4188 |
| Cairo | -61.4714 | 66.7377 | 0.36 | 66.3777 |
| Longonot | -18.0488 | 23.3151 | 0.36 | 22.9551 |
| Port Louis | -34.3061 | 39.5724 | 0.36 | 39.2124 |
| Praia | -40.0548 | 44.6911 | 0.36 | 44.3311 |

close. Additionally, we computed the CNR for the stations for link unavailability time of 0.001%, 0.1% and 1%. The complete results for each of these unavailability times are depicted in **Error! Reference source not found.** and

.

| Ground Station | Rainfall Attenuation (dB) | | Overestimation Percentage |
|---|---|---|---|
| | TRMM | GPM | |
| Abuja | 10.5587 | 10.3059 | 102.3942 |
| Hartbeesthoek | 22.7269 | 22.7947 | 99.7017 |
| Cairo | -7.8440 | -13.2802 | 100.6930 |
| Longonot | 15.5252 | 16.3896 | 105.5677 |
| Port Louis | -0.3620 | 0.7753 | 101.373 |
| Praia | -1.7871 | -1.3956 | 106.6934 |

*Table 5 Estimation comparison of TRMM and GPM results for the six candidate ground stations.*

| Ground Station | Available link (dB) for different unavailability times | | | | |
|---|---|---|---|---|---|
| | 1% | 0.5% | 0.1% | 0.01% | 0.001% |
| Abuja | 9.5255 | 1.718 | 3.6989 | -7.5994 | -5.1025 |
| Hartbeesthoek | -5.5514 | -13.2145 | -10.3438 | -20.0882 | -16.0089 |
| Cairo | 37.9994 | 29.1518 | 29.4063 | 22.1473 | 16.3977 |
| Longonot | 2.1810 | -5.3206 | -2.682 | -13.6831 | -10.6924 |
| Port Louis | 21.0312 | 12.7648 | 14.0458 | 3.4818 | 3.6307 |
| Praia | 23.6519 | 15.3909 | 16.519 | 4.1021 | 5.4907 |

*Table 6 Available link results for the six candidate stations for unavailability time of 1%, 0.5%, 0.1%, 0.01% and 0.001% and uplink centre frequency of 28.5GHz*

The correlation in estimation percentages is not consistent for all the stations. For instance, at Hartbeesthoek, the GPM results underestimate the attenuation by about 0.3%. The inconsistency can be attributed to the different times of the data used. The TRMM covers 2005-2014 while the TRMM, 2015-2020. Therefore, no meaningful comparison can be done for the two sets of the results. Furthermore, there are improvements in GPM measurements as highlighted in the data section. Due to those reasons, the GPM results are adopted henceforth to calculate the link margin for the stations in this paper.

Since the CNR for 0.01% unavailability time for average year is a small and ideal value, calculations were done for 0.5% as per Telesat's filing using GPM results [15]. The results are presented in **Error! Reference source not found.**.

The CNR results for 0.5% unavailability time reveals that only Hartbeesthoek and Longonot link do not

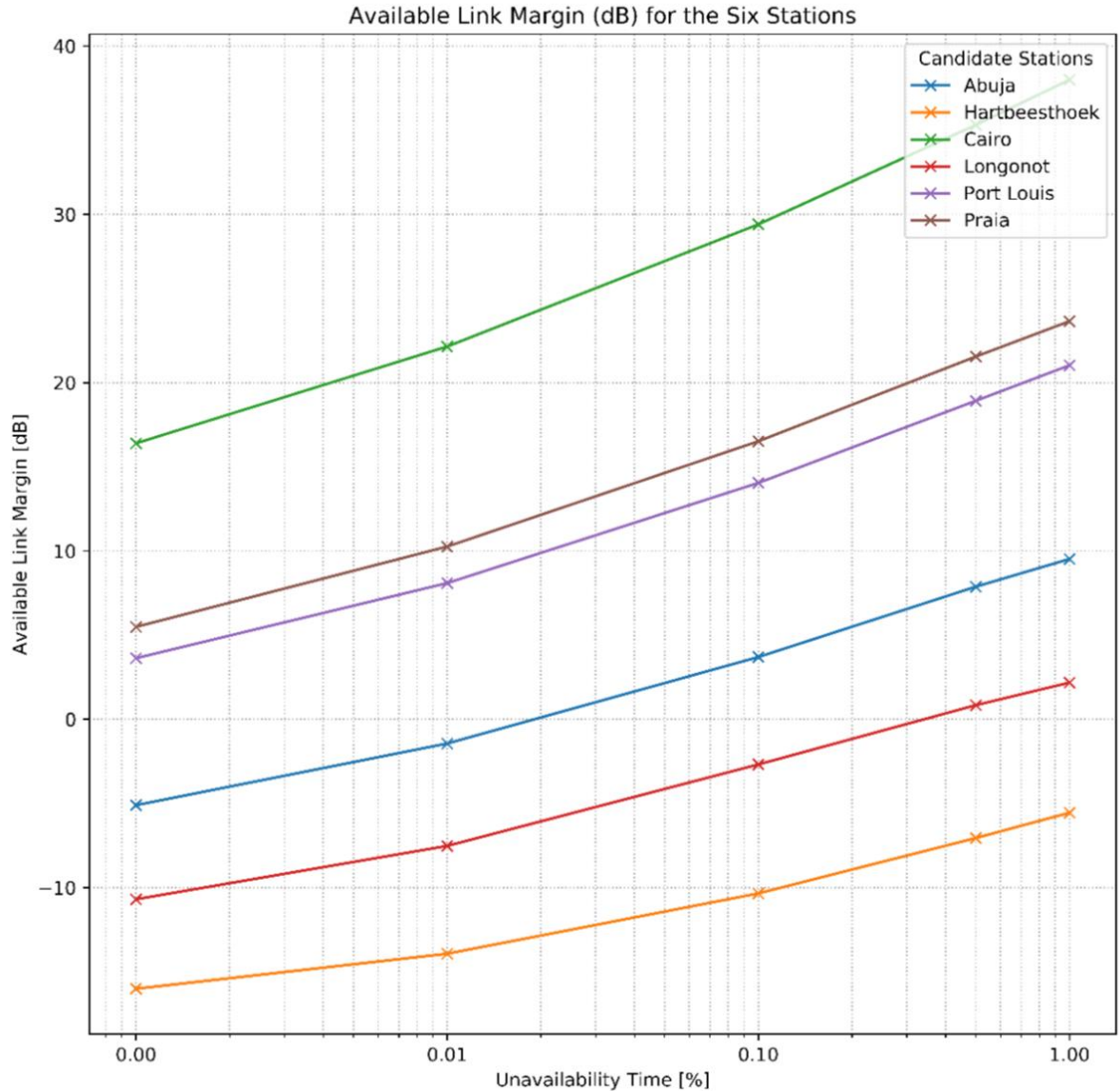

*Figure 6 Available link (dB) results for the six candidate stations for unavailability time of 1%, 0.5%, 0.1%, 0.01% and 0.001% and uplink centre frequency of 28.5GHz*

. investigated their response to lower frequency C-band

| Ground Station | Rainfall Attenuation (dB) | CNR (dB) | Margin (dB) | Available Margin (dB) |
|---|---|---|---|---|
| Abuja | -24.1363 | 29.4026 | 0.36 | 29.0426 |
| Hartbeesthoek | -13.5125 | 18.7788 | 0.36 | 18.4188 |
| Cairo | -61.4714 | 66.7377 | 0.36 | 66.3777 |
| Longonot | -18.0488 | 23.3151 | 0.36 | 22.9551 |
| Port Louis | -34.3061 | 39.5724 | 0.36 | 39.2124 |
| Praia | -40.0548 | 44.6911 | 0.36 | 44.3311 |

*Table 7 Results of the rainfall attenuation, CNR and available link margin based on GPM data for all the six candidate ground stations for C band Uplink frequency (6GHz) and 0.01% unavailability time*

Due to the non-closure of Longonot (for 0.01%) and Hartbeesthoek (for 0.01% and 0.5%) links, we then (4-8GHz) that is dominantly used in earth-space communication for 0.01% unavailability time. **Error! Reference source not found.** shows the CNR and available link margin for the six stations at 6GHz centre frequency. At C-band, the link for all stations close.

IV. DISCUSSION

Based on ITU-R P.837-7 rain intensity results, South Africa's, Hartbeesthoek and Kenya's, Longonot have the highest rainfall attenuation of $\approx 49dB$ and $\approx 40dB$. On the contrary, Mauritius', Port Louis and Cape Verde's recorded the least attenuation of $\approx 28dB$. Generally, none of the candidate stations could have their links close when the results of ITU-R P.837-7 are used. This implies that all the transmitted power from the ground station would not reach the satellite when the rain attenuation results are assumed. That prompts the optimization of system parameters such as antenna diameter, EIRP, noise power among other factors. However, that increases the mission cost. This partly explains why other global LEO constellation companies such as OneWeb and SpaceX opted for higher elevation angles of 55º and 40º respectively to overcome the atmospheric attenuation in tropical and equatorial area [15].

The link performance results of the ground stations due to rainfall attenuation is also presented based on the rain rate data from TRMM. Unlike the case of ITU-R P.837-7, Port Louis, Praia and Cairo links closes with 0.01% unavailability time. They can be used for high data rate communication applications and will only be unavailable for $\approx 53 minutes$ in a year. This is a great improvement from the results obtained with ITU-R P.837-7 values as it shows that extra system modifications are unnecessary for the links to close. Setting up a LEO communication ground system in these regions sets a lower engineering threshold that reduces the total mission cost. Abuja station also improved and minimal modifications can be applied to close its link. It is also available for high data rate communication.

Using GPM data, Kenya's, Longonot and South Africa's, Hartbeesthoek stations do not surpass the link margin threshold for 0.01% unavailability time. It implies that, with Telesat's system requirement specifications as in Table 1, these two stations are unable to communicate at all with the satellites. This is attributed to high rainfall attenuation results (16.3896dB and 22.7947dB respectively) due to their geographical location. They are in regions with high rain intensity. The other stations surpass the link margin and can be used for high data rate communications as anticipated in the emerging LEO systems such as Telesat, Starlink and OneWeb [15] [22]. Coincidentally, the two stations have the highest altitudes of 1385m and 1715m, respectively. This could have contributed to the non-closure of the links since it determines the slant path length in the ITU-R P.837-7. Additionally, these stations are available for low frequency C-band (6GHz) transmission especially for telemetry, backhaul and voice call applications as confirmed by the results of **Error! Reference source not found.**. The 6GHz centre frequency is resilient to rain attenuation since degradation of radio signals starts at frequencies above 7GHz. Therefore, future LEO communication missions across Africa should avoid regions experiencing high rainfall such as the Congo Basin and greater central African region unless they are used for low data rate transmission and backhaul stations.

Cairo records the best CNR due to its least rainfall attenuation, followed by Praia, Port Louis and Abuja. These stations recorded the least rain intensity thus contributing to high CNR values. They are the ideal stations for supporting LEO communication systems with a high QoS of 0.01% unavailability time. However, as per Telesat's filing, the unavailability time is set to 0.5% implying that the company intends to only have disrupted service for a cumulative duration of 44 hours in a year. Therefore, setting the unavailability time at 0.01% (53minutes) for its service is unrealistic. The cost benefit of designing a satellite communication system that is only unavailable for a cumulative time of 53 minutes is high thus prompting the need for setting bigger time of 0.5%. Even so, with this ideal margin, Cairo, Praia and Port Louis surpass the link margin thus reaffirming their ability to be a gateway station for a LEO communication system such as Telesat over Africa.

At 0.5% unavailability time, Abuja candidate station surpasses the link and can therefore be used as a gateway hub too. However, Longonot misses the margin with -3.1208dB and can be forced to close through engineering modification such as increasing the antenna aperture and reduction of the noise temperature. The current Telesat antenna size adopted in this study is 3.5m. increasing the size will significantly raise the Effective Isotropic radiated power (EIRP) of the gateway station. However, that prompts for modification of the satellite receiver system to handle the rise in power received. Furthermore, it affects the system noise generated by the satellite's electronics [20]. That calls for optimal balance between increasing the EIRP and reducing the noise generated.

Unlike Kenya, Hartbeesthoek do not achieve the link margin prompting the need for calculation for 0.001%, 0.1% and 1% unavailability time. Just like the 0.01%, 0.001% is unrealistic as it equates to a cumulative time of 5minutes service disruption in a year. Nevertheless,

Cairo, Praia and Port Louis still meets this high threshold. These stations provide the best terminal site for a system such as Telesat. On the other hand, the 1% unavailability time is a huge number corresponding to a cumulative period of about four days in a year. This is not feasible for broadband internet service provider. However, the 0.1% unavailability time is more desirable. Cairo, Praia and Port Louis stations satisfy the link margin requirements at that threshold.

When the performance of the Longonot and Hartbeesthoek stations were tested for low frequency C-band (4-8GHz) at 6GHz centre frequency, the results were different. Even at the ideal 0.01% unavailability time, Hartbeesthoek and Longonot stations close without requiring further modifications. This is expected since rainfall attenuation in the tropics and equatorial region where these stations are located are only severe at frequencies above 7GHz [23] [10]. At that frequency, the two stations can be used for low-data rate communications, backhaul or tracking, telemetry and command (TTC) functions.

The link performance results based on rain rate data from GPM shows that Cairo, Praia and Port Louis are the best gateway station for new generation LEO communication satellites such as Telesat. They can achieve high QoS even at ideal link margins of 0.001%, 0.01% and 0.1%. at Telesat's recommended 0.5%, Abuja can also serve as a ground terminal for high link data rate station. However, the Longonot and Hartbeesthoek stations require significant modifications for high QoS at the recommended unavailability time of 0.5% a year. For LEO communication missions, ground stations should be located at upper latitudes (>10°N) where rain attenuation is lower as is illustrated by Cairo, Praia and Abuja. Additionally, the location of stations should be in low altitudes as the link margin of Longonot and Hartbeesthoek has been low even though their rain rate level does not vary significantly compared to Abuja that was located in a lower altitude. To guarantee reliable satellite communication with low unavailability time (<0.01%), transmission antenna aperture sizes should be at least above 3.5m to increase the EIRP and subsequent CNR on the space segment.

## V. CONCLUSIONS

The rain attenuation for six candidate ground stations across Africa (Cairo, Praia, Port Louis, Abuja, Longonot and Hartbeesthoek) is computed using the ITU-R P.618-8 model and GPM satellite rain rate data. From the attenuation results, the CNR for each of the stations at 0.001%-1% unavailability time is calculated to determine the best candidate station for a LEO communication system such as Telesat. The results reveal that the candidate stations located in low rain intensity regions (Cairo, Praia and Port Louis) are the most suitable terminals at 0.001%, 0.01% and 0.5% unavailability time. Notably, Port Louis records high CNR although its rain intensity rates are higher. Abuja candidate station performs well when unavailability time is set to Telesat's recommended 0.5% hence also available as a potential gateway hub. However, Longonot and Hartbeesthoek are the worst performing stations to be a gateway unless used for low-data rate transmission in C-band (6GHz centre frequency) transmission.

The rain rates results were based on merged GPM satellite data and did not consider the point rain gauge measurement of each of the candidate ground stations. In future, using point rain gauge measurements for each of these stations may yield a higher accurate attenuation results to be used for the link performance analysis of LEO communication satellite ground terminals. Secondly, the CNR calculation did not take care of other system losses such as polarizing, mismatch, depointing, scintillation, cloud and gaseous attenuation. Moreover, the system parameters used are based on literature and those revealed by Telesat to the public. Some crucial system parameters are commercial secret for the company thus limiting the conclusivity of the final results. In future, we anticipate conducting a complete analysis of cloud, gaseous and scintillation attenuation to fully characterize and establish the link margin for these proposed stations in future.